# AutoSpearman: Automatically Mitigating Correlated Software Metrics for Interpreting Defect Models

Jirayus Jiarpakdee, Chakkrit Tantithamthavorn, Christoph Treude
School of Computer Science,
The University of Adelaide, Australia.
firstname.lastname@adelaide.edu.au

*Abstract*—The interpretation of defect models heavily relies on software metrics that are used to construct them. However, such software metrics are often correlated in defect models. Prior work often uses feature selection techniques to remove correlated metrics in order to improve the performance of defect models. Yet, the interpretation of defect models may be misleading if feature selection techniques produce subsets of inconsistent and correlated metrics. In this paper, we investigate the consistency and correlation of the subsets of metrics that are produced by nine commonly-used feature selection techniques. Through a case study of 13 publicly-available defect datasets, we find that feature selection techniques produce inconsistent subsets of metrics and do not mitigate correlated metrics, suggesting that feature selection techniques should not be used and correlation analyses must be applied when the goal is model interpretation. Since correlation analyses often involve manual selection of metrics by a domain expert, we introduce `AutoSpearman`, an automated metric selection approach based on correlation analyses. Our evaluation indicates that `AutoSpearman` yields the highest consistency of subsets of metrics among training samples and mitigates correlated metrics, while impacting model performance by 1-2%pts. Thus, to automatically mitigate correlated metrics when interpreting defect models, we recommend future studies use `AutoSpearman` in lieu of commonly-used feature selection techniques.

## I. INTRODUCTION

Defect models are statistical or machine learning models that are used to investigate the impact of software metrics (e.g., lines of code) on defect-proneness and to identify defect-prone software modules. The interpretation of defect models is used to formulate empirical theories related to software quality, which are essential to chart quality improvement and maintenance plans.

The interpretation of defect models heavily relies on the software metrics that are used to construct them. However, software metrics often have strong correlation among themseleves [25, 35, 72, 74, 85]. Prior work [35] finds that many metrics in defect datasets are *correlated* (e.g., the `branch_count` metric is linearly proportional to the `decision_count` metric in some NASA datasets). Thus, correlated metrics often have a negative impact on the interpretation of defect models [4, 34, 68, 74, 85]. For example, the most important characteristics of defective modules that are derived from defect models may be incorrect, leading to misleading quality improvement and maintenance plans.

To address these concerns, feature selection techniques are often applied to remove correlated metrics [3, 8, 12, 18, 38, 54, 58, 67]. In practice, *feature selection techniques should be applied only on training samples* to avoid producing optimistically biased performance estimates and interpretation [45]. Feature selection techniques may produce different subsets of metrics for each training sample that is randomly generated from model validation techniques. Such inconsistency of subsets of metrics often leads to different interpretation among training samples. Yet, little is known about whether feature selection techniques consistently produce the same subset of metrics among different training samples and among these feature selection techniques in defect datasets.

In addition, Jiarpakdee *et al.* [34] raise concerns that defect models that are constructed with such correlated metrics may produce different interpretation by simply rearranging a model specification (e.g., from $y \sim m_1 + m_2$ to $y \sim m_2 + m_1$ if $m_1$ and $m_2$ are correlated). Yet, little is known about whether feature selection techniques mitigate correlated metrics in defect datasets.

In this paper, we set out to investigate the consistency and correlation of subsets of metrics that are produced by feature selection techniques. We apply nine feature selection techniques from two families that are commonly-used in the defect prediction domain, i.e., four filter-based and five wrapper-based feature selection techniques. Through a case study of 13 publicly-available defect datasets of systems that span both proprietary and open source domains, we address the following two research questions:

**(RQ1) Do feature selection techniques consistently produce the same subset of metrics in defect datasets?**
Feature selection techniques produce inconsistent subsets of metrics. When applying the studied feature selection techniques to different training samples from the same dataset, we find that only 6-41% of the metrics are consistently selected. On the other hand, when applying the studied feature selection techniques to the same training sample, only 0-15% of the metrics are consistently selected.

**(RQ2) Do feature selection techniques mitigate correlated metrics in defect datasets?**

Feature selection techniques do not mitigate correlated metrics. The studied feature selection techniques produce up to 100% of subsets of metrics with collinearity and multicollinearity, suggesting that correlation analyses (e.g., a Spearman rank correlation test and a Variance Inflation Factor analysis (VIF)) should be applied to mitigate correlated metrics.

Since our findings suggest that commonly-used feature selection techniques should not be used to mitigate correlated metrics and correlation analyses must be applied, we then further investigate if such correlation analyses increase the consistency of subsets of metrics and impact the performance of defect models. However, correlation analyses (i.e., a Spearman rank correlation test and a Variance Inflation Factor test) involve manual metric selection from a domain expert and thus the automation of software analytics is limited in practice [14].

To address these concerns, we introduce `AutoSpearman`, an automated metric selection approach based on correlation analyses for statistical inference. `AutoSpearman` uses the Spearman rank correlation test and the Variance Inflation Factor analysis to identify and mitigate correlated metrics. To evaluate the consistency of subsets of metrics that are produced by `AutoSpearman`, and its impact on the performance of defect models, we address the following two research questions:

**(RQ3) What is the consistency of subsets of metrics in defect datasets that are produced by `AutoSpearman`?**

`AutoSpearman` yields the highest consistency of subsets of metrics among different training samples when comparing to all studied feature selection techniques. When applying `AutoSpearman` to different training samples from the same dataset, we find that, at the median, 69% of the metrics are consistently selected, which leads to improvements of up to 86% from the studied commonly-used feature selection techniques.

**(RQ4) What is the impact of `AutoSpearman` on the performance of defect models?**

`AutoSpearman` only impacts the performance of defect models by 1-2%pts for AUC, F-measure, and MCC measures, respectively.

Our evaluation indicates that `AutoSpearman` yields the highest consistency of subsets of metrics among training samples and mitigates correlated metrics with little impact on model performance. Thus, to automatically mitigate correlated metrics when interpreting defect models, we recommend future studies use `AutoSpearman` in lieu of commonly-used feature selection techniques.

**Novelty and Contributions**. To the best of our knowledge, this is the first paper to:

(1) Investigate the consistency of subsets of metrics that are produced by feature selection techniques (RQ1).
(2) Investigate the correlation of subsets of metrics that are produced by feature selection techniques (RQ2).
(3) Introduce `AutoSpearman`, an automated metric selection approach based on correlation analyses (Section V). Unlike manual metric selection by a domain expert which may be subjective, `AutoSpearman` uses a new criterion for automated metric selection to **select one metric of a group of the highest correlated metrics that shares the least correlation with other metrics that are not in that group**. Unlike commonly-used feature selection techniques which consider the relationship between metrics and the outcome (e.g., wrapper-based feature selection), `AutoSpearman` is performed in an unsupervised fashion (i.e., outcome labels are not required) to avoid producing optimistically biased performance estimates and interpretation. We also provide an implementation of `AutoSpearman` as an R package [2].
(4) Investigate the consistency of subsets of metrics that are produced by `AutoSpearman` (RQ3).
(5) Investigate the performance of defect models that are constructed from subsets of metrics when applying `AutoSpearman` (RQ4).

**Paper Organisation**. Section II describes the studied feature selection techniques. Section III discusses the criteria for selecting the studied datasets. Section IV presents the motivation, approach, and results with respect to research questions 1 and 2. Section V introduces a new automated metric selection approach based on correlation analyses, while Section VI presents the motivation, approach, and evaluation results of `AutoSpearman` with respect to research questions 3 and 4. Section VII elaborates on the threats to the validity of our study. Finally, Section VIII draws conclusions.

## II. FEATURE SELECTION TECHNIQUES

Feature selection is a data preprocessing technique for selecting a subset of the best software metrics prior to constructing a defect model. Feature selection has been widely used in software engineering to remove *irrelevant* metrics (i.e., metrics that do not share a strong relationship with the outcome) [52, 64] and *correlated* metrics (i.e., metrics that share a strong correlation with one or more metrics) [25, 35, 74, 85]—assuming that the removal of such metrics will produce interpretable defect models without impacting their performance. Such interpretable defect models allow us to derive more actionable insights for researchers and practitioners [70]. Nevertheless, prior work points out that metrics selected by feature selection techniques are often correlated [84].

Indeed, many research efforts have shown the benefits of applying feature selection techniques to defect prediction models [24, 46, 83]. For example, Ghotra *et al.* [24], Xu *et al.* [83], and Lu *et al.* [46] investigate the impact of feature selection techniques on the performance of defect models. Yet, no prior work has investigated the impact of feature selection techniques when interpreting defect models.

There is a plethora of feature selection techniques that can be applied [26], e.g., filter-based, wrapper-based, and embedded-based families. Since it is impractical to study all of these techniques, we would like to select a manageable set of feature selection techniques for our study. Similar to Ghotra *et al.* [24], we select two commonly-used families of feature selection techniques, i.e., filter-based feature selection techniques and wrapper-based feature selection techniques. Thus, embedded-based feature selection techniques are excluded from our analysis, as they are rarely explored in software engineering. Below, we provide the description of each studied feature selection technique and their detailed implementation.

Table I: A summary of the detailed implementation for the nine studied feature selection techniques.

| Type | Technique | R Package | R Function | Abbreviation |
|---|---|---|---|---|
| Filter-based Feature Selection Techniques | Correlation-based | FSelector [62] | `cfs(class~metrics, dataset)` | CFS |
| | Information Gain | | `information.gain(class~metrics, dataset)` | IG |
| | Chi-Squared-based ($\chi^2$) | | `chi.squared(class~metrics, dataset)` | Chisq |
| | Consistency-based | | `consistency(class~metrics, dataset)` | CON |
| Wrapper-based Feature Selection Techniques | Recursive Feature Elimination (Logistic Regression) | caret [43] | `lrFuncs.AUC = lrFuncs`<br>`lrFuncs.AUC$summary = twoClassSummary`<br>`control = rfeControl(functions = lrFuncs.AUC, method = "boot", number = iterations)`<br>`rfe(x = dataset[, metrics], y = dataset[, class], rfeControl = control, metric = "ROC")` | RFE-LR |
| | Recursive Feature Elimination (Random Forest) | | `rfFuncs.AUC = rfFuncs`<br>`rfFuncs.AUC$summary = twoClassSummary`<br>`control = rfeControl(functions = rfFuncs.AUC, method = "boot", number = iterations)`<br>`rfe(x = dataset[, metrics], y = dataset[, class], rfeControl = control, metric = "ROC")` | RFE-RF |
| | Stepwise Regression (Forward Direction) | stats [77] | `null.model = glm(class~1, data = dataset, family = binomial())`<br>`full.model = glm(class~metrics, data = dataset, family = binomial())`<br>`step(null.model, scope = list(upper = full.model), data = dataset, direction = "fwd")` | Step-FWD |
| | Stepwise Regression (Backward Direction) | | `full.model = glm(class~metrics, data = dataset, family = binomial())`<br>`step(full.model, data = dataset, direction = "bwd")` | Step-BWD |
| | Stepwise Regression (Both Directions) | | `null.model = glm(class~1, data = dataset, family = binomial())`<br>`full.model = glm(class~metrics, data = dataset, family = binomial())`<br>`step(null.model, scope = list(upper = full.model), data = dataset, direction = "both")` | Step-BOTH |

### A. Filter-based feature selection techniques

Filter-based feature selection techniques search for the best subset of metrics according to an evaluation criterion regardless of model construction. Since constructing models is not required, the use of filter-based feature selection techniques is considered low cost and widely used in the defect prediction literature [3, 8, 18, 38, 54, 58]. There are many variants of filter-based feature selection techniques, which we describe below.

**Correlation-based feature selection** [27] searches for the best subset of metrics that shares the strongest relationship with the outcome, while having a low correlation among themselves.

**Information gain feature selection** [55] ranks metrics according to the information gain with respect to the outcome. The information gain is measured by how much information of the outcome is provided by a metric.

**Chi-Squared-based feature selection** [48] assesses the importance of metrics with the $\chi^2$ statistic which is a non-parametric statistical test of independence.

**Consistency-based feature selection** [13] uses the consistency measure (i.e., inconsistency rate) to evaluate a subset of metrics. The technique finds the optimal subset of metrics whose inconsistency rate approximates the inconsistency rate of all metrics.

Table I summarises the detailed implementation of the studied filter-based feature selection techniques.

### B. Wrapper-based Feature Selection Techniques

Wrapper-based feature selection techniques [36, 40] use classification techniques to assess each subset of metrics and find the best subset of metrics according to an evaluation criterion. Wrapper-based feature selection is made up of three steps, which we describe below.

*(Step 1) Generate a subset of metrics.* Since it is impossible to evaluate all possible subsets of metrics, wrapper-based feature selection often uses search techniques (e.g., best first, greedy hill climbing) to generate candidate subsets of metrics for evaluation.

*(Step 2) Construct a classifier using a subset of metrics with a predetermined classification technique.* Wrapper-based feature selection constructs a classification model using a candidate subset of metrics for a given classification technique (e.g., logistic regression and random forest).

*(Step 3) Evaluate the classifier according to a given evaluation criterion.* Once the classifier is constructed, wrapper-based feature selection evaluates the classifier using a given evaluation criterion (e.g., Akaike Information Criterion).

For each candidate subset of metrics, wrapper-based feature selection repeats Steps 2 and 3 in order to find the best subset of metrics according to the evaluation criterion. Finally, it provides the best subset of metrics that yields the highest performance according to the evaluation criterion.

In this study, we select two commonly-used variants of wrapper-based feature selection techniques, which we describe below.

**Recursive Feature Elimination** (RFE) [26] searches for the best subset of metrics by recursively eliminating the least important metric. First, RFE constructs a model using all metrics and ranks metrics according to their importance score (e.g., Breiman's Variable Importance for random forest). In each iteration, RFE excludes the least important metric and reconstructs a model. Finally, RFE provides the subset of metrics which yields the best performance according to an evaluation criterion (e.g., AUC). In our study, we select the AUC measure since it measures the discriminatory power of models, as suggested by recent research [23, 44, 60, 71]. We use the implementation of the recursive feature elimination using the `rfe` function as provided by the `caret` R package [43].

Table II: A statistical summary of the studied datasets.

| Project | Dataset | Modules | Metrics | Defective Ratio | EPV | $AUC_{LR}$ | $AUC_{RF}$ |
|---|---|---|---|---|---|---|---|
| Apache | Xalan 2.6 | 885 | 20 | 46 | 21 | 0.79 | 0.85 |
| Eclipse | Debug 3.4 | 1,065 | 17 | 25 | 15 | 0.72 | 0.81 |
| | JDT | 997 | 15 | 21 | 14 | 0.81 | 0.82 |
| | Mylyn | 1,862 | 15 | 13 | 16 | 0.78 | 0.74 |
| | PDE | 1,497 | 15 | 14 | 14 | 0.72 | 0.72 |
| | Platform 2 | 6,729 | 32 | 14 | 30 | 0.82 | 0.84 |
| | Platform 2.1 | 7,888 | 32 | 11 | 27 | 0.77 | 0.78 |
| | Platform 3 | 10,593 | 32 | 15 | 49 | 0.79 | 0.81 |
| | SWT 3.4 | 1,485 | 17 | 44 | 38 | 0.87 | 0.97 |
| Proprietary | Prop 1 | 18,471 | 20 | 15 | 137 | 0.75 | 0.79 |
| | Prop 2 | 23,014 | 20 | 11 | 122 | 0.71 | 0.82 |
| | Prop 4 | 8,718 | 20 | 10 | 42 | 0.74 | 0.72 |
| | Prop 5 | 8,516 | 20 | 15 | 65 | 0.7 | 0.71 |

**Stepwise regression** [10] finds the best subset of metrics by individually assessing each metric and adding (or removing) a metric if it improves an evaluation criterion (e.g., Akaike Information Criterion). The process is repeated until there is no improvement from adding or removing a metric. In this paper, we study three directions (i.e., forward, backward, and both directions) of the stepwise regression technique. We use the implementation of stepwise regression using the `step` function as provided by the `stats` R package [77].

## III. STUDIED DATASETS

In selecting the studied datasets, we identify four important criteria that need to be satisfied:

**Criterion 1—Publicly-available defect datasets.** Prior work raises concerns about the replicability of software engineering studies [61]. In order to foster future replication of our work, we focus on publicly-available defect datasets.

**Criterion 2—Datasets that are reliable and of high quality.** Defect models rely greatly on the quality of the datasets that are used to construct them. Shepperd *et al.* [64] raise concerns related to data quality in the NASA datasets. Furthermore, Petrić *et al.* [59] show that problematic data remain in the cleaned NASA datasets. Thus, the quality of the NASA datasets is questionable. To ensure that the studied datasets are reliable and of high quality, we exclude the NASA datasets from our study.

**Criterion 3—Datasets that produce non-overly optimistic model performance.** Classification techniques that are trained on imbalanced data often favour the majority class. When defective modules are the majority class, defect models are likely to produce overly optimistic performance estimates. Thus, we exclude datasets that have a defective ratio above 50%.

**Criterion 4—Datasets where we can accurately derive interpretations.** Analysts would only consider models that fit the data well (i.e., AUC > 0.7) and are stable (i.e., EPV > 10) [75]. Hence, we only focus on datasets that produce such accurate and stable models.

To satisfy criterion 1, similar to prior work [73], we begin our study using a collection of the 101 publicly-available defect datasets that are collected from 5 different corpora, i.e., 76 datasets from the Tera-PROMISE Repository [53], 12 clean NASA datasets as provided by Shepperd *et al.* [64], 5 datasets as provided by Kim *et al.* [39, 82], 5 datasets as provided by D'Ambros *et al.* [11, 12], and 3 datasets as provided by Zimmermann *et al.* [87]. To satisfy criterion 2, we exclude 12 datasets where their data quality is questionable. To satisfy criterion 3, we exclude 17 datasets that have a defective ratio above 50%. Finally, to satisfy criterion 4, we exclude 59 datasets which have an EPV value below 10 and produce models with an AUC value below 0.7. Hence, we focus on 13 datasets of systems that span across proprietary and open-source systems. Table II shows a statistical summary of the 13 studied datasets.

## IV. CASE STUDY RESULTS

In this section, we present the motivation, approach, and results with respect to the first two research questions.

*(RQ1) Do feature selection techniques consistently produce the same subset of metrics in defect datasets?*

**Motivation**. In practice, *feature selection techniques should only be applied on training samples* because of the unavailability of defect labels in testing samples. Since training samples are often randomly generated from model validation techniques (e.g., out-of-sample bootstrap validation or 10-folds cross-validation), feature selection techniques may produce different subsets of metrics for each training sample. Different subsets of metrics among training samples may pose a critical threat to validity when analysing and identifying the most important metrics. For example, prior work often applies post-hoc multiple comparison analyses (e.g., a Scott-Knott test) on the distributions of importance scores to identify statistical distinct ranks of the most important metrics [34, 71, 80]. Thus, such post-hoc analyses cannot be applied when feature selection techniques produce different subsets of metrics. Furthermore, the conclusions of prior work often rely on one feature selection technique [3, 12, 18, 38, 54] which may pose a threat to the construct validity, i.e., conclusions may not hold true if another feature selection technique is applied. Nevertheless, little is known about whether feature selection techniques produce the same subset of metrics among training samples and among feature selection techniques in defect datasets.

**Approach**. To address RQ1, we investigate two perspectives of the consistency of the subsets of metrics, i.e., the consistency among training samples and the consistency among feature selection techniques. We first generate training samples. Then, we apply feature selection techniques on each training sample. Finally, we analyse the consistency of subsets of metrics that are produced by the studied feature selection techniques. We describe each step below.

*(Step 1) Generate training samples.* To generate training samples, we use the out-of-sample bootstrap validation technique that (1) leverages aspects of statistical inference [17, 21, 31]; and (2) produces the least bias and variance of performance estimates for defect prediction [75]. We randomly generate a bootstrap sample of size $N$ with replacement from an original dataset, where $N$ is the size of the original dataset. On average, 36.8% of the original dataset will not be selected, since a bootstrap sample is selected with replacement [17]. We repeat the out-of-sample bootstrap process 100 times.

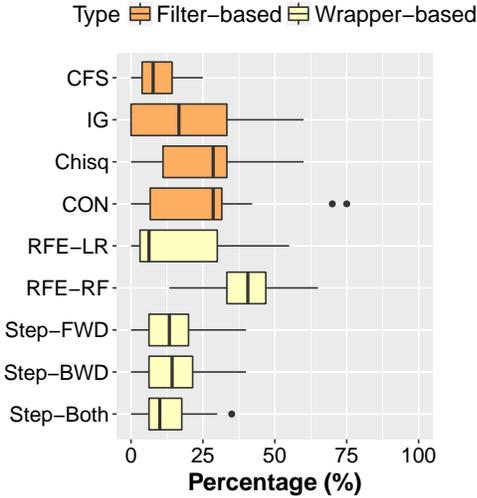

Figure 1: The percentage of metrics that are consistently selected when applying feature selection techniques to different training samples from the same dataset.

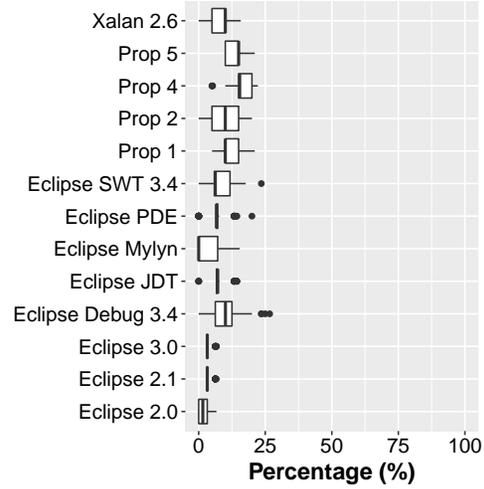

Figure 2: The percentage of metrics that are consistently selected when applying feature selection techniques to the same training sample for all defect datasets.

*(Step 2) Apply feature selection techniques.* We only apply feature selection techniques on *a training sample*, instead of the whole dataset in order to avoid producing optimistically biased performance estimates and interpretation [45]. For each feature selection technique, we use the implementation as shown in Table I. The subsets of metrics that are produced by each studied feature selection technique for all studied datasets are available in the online appendix [1].

*(Step 3) Analyse the consistency of subsets of metrics among different training samples.* We start from the subsets of metrics for all of the 100 training samples for each feature selection technique of each studied dataset. Ideally, each feature selection technique should produce the same subset of metrics for all of the 100 training samples. We compute the consistency as a percentage of the unique metrics that consistently appeared among all of the 100 training samples and all of the unique metrics for all training samples (i.e., $\frac{|S_{\mathrm{FS}_i\mathrm{TS}_1} \cap S_{\mathrm{FS}_i\mathrm{TS}_2} ... \cap S_{\mathrm{FS}_i\mathrm{TS}_{100}}|}{|S_{\mathrm{FS}_i\mathrm{TS}_1} \cup S_{\mathrm{FS}_i\mathrm{TS}_2} ... \cup S_{\mathrm{FS}_i\mathrm{TS}_{100}}|}$, where $S_{\mathrm{FS}_i\mathrm{TS}_j}$ is a subset of metrics that is produced by a feature selection ($\mathrm{FS}_i$) when applied on a training sample ($\mathrm{TS}_j$)). We present the consistency percentage using boxplots in Figure 1.

*(Step 4) Analyse the consistency of subsets of metrics among different feature selection techniques.* We start from the subsets of metrics that are produced by all studied feature selection techniques for each training sample of each studied dataset. Ideally, these techniques should consistently produce the same subset of metrics. We measure the consistency as a percentage of the unique metrics that consistently appeared among all of the studied feature selection techniques and all of the unique metrics for all studied feature selection techniques (i.e., $\frac{|S_{\mathrm{FS}_1\mathrm{TS}_j} \cap S_{\mathrm{FS}_2\mathrm{TS}_j} ... \cap S_{\mathrm{FS}_9\mathrm{TS}_j}|}{|S_{\mathrm{FS}_1\mathrm{TS}_j} \cup S_{\mathrm{FS}_2\mathrm{TS}_j} ... \cup S_{\mathrm{FS}_9\mathrm{TS}_j}|}$, where $S_{\mathrm{FS}_i\mathrm{TS}_j}$ is a subset of metrics that is produced by a feature selection ($\mathrm{FS}_i$) when applied on a training sample ($\mathrm{TS}_j$)). We present the consistency percentage using boxplots in Figure 2.

**Results**. **When applying the studied feature selection techniques to different training samples from the same dataset, only 6-41% of the metrics are consistently selected.** Figure 1 shows the percentage of metrics that are consistently selected when applying feature selection techniques to different training samples. We find that, at the median, only 8%, 17%, 30%, 30%, 6%, 41%, 13%, 14%, and 10% of the metrics are consistently selected when applying CFS, IG, Chisq, CON, RFE-LR, RFE-RF, Step-FWD, Step-BWD, and Step-BOTH to different training samples. Although the training samples are drawn from the same original dataset, none of the studied feature selection techniques consistently produce the same subset of metrics. This finding suggests that the randomisation of training samples could produce defect models that are constructed from different subsets of metrics even if the training samples are drawn from the same dataset. Such inconsistency of subsets of metrics restricts an application of post-hoc multiple comparison analyses (e.g., a Scott-Knott test) to identify the most important metrics when interpreting defect models.

**When applying the studied feature selection techniques to the same training sample, only 0-15% of the metrics are consistently selected.** Figure 2 shows the percentage of metrics that are consistently selected when applying feature selection techniques on each training sample of each defect dataset. We observe that, at the median, only 0-15% of the

metrics are consistently selected among the studied feature selection techniques, suggesting that the conclusions (e.g., the impact of metrics on an outcome) of prior work may not hold true if other feature selection techniques are applied.

The low consistency among feature selection techniques has to do with (1) the variants of the evaluation criteria of filter-based feature selection techniques; and (2) the variants of the classification techniques of wrapper-based feature selection techniques. For filter-based feature selection techniques, the best metrics according to an evaluation criterion (e.g., information gain) might not be the best for the other criteria (e.g., $\chi^2$ statistic). For wrapper-based feature selection techniques, the best metrics according to a classification technique (e.g., logistic regression) might not be the best for the other classification techniques (e.g., random forest). These observations suggest that *future studies must report the settings of the used feature selection techniques* (e.g., the used evaluation criterion of filter-based feature selection teachniques and the used classification technique for wrapper-based feature selection techniques).

> *Feature selection techniques produce inconsistent subsets of metrics. When applying the studied feature selection techniques to different training samples from the same dataset, we find that only 6-41% of the metrics are consistently selected. On the other hand, when applying the studied feature selection techniques to the same training sample, only 0-15% of the metrics are consistently selected.*

*(RQ2) Do feature selection techniques mitigate correlated metrics in defect datasets?*

**Motivation**. The conclusions of prior defect studies rely on the usage of built-in interpretation techniques of classification techniques (e.g., ANOVA for logistic regression, and Breiman's Variable Importance for random forest). However, recent work points out that such interpretation techniques are sensitive to correlated metrics [4, 34, 68, 74, 85]. For example, Jiarpakdee et al. [34] show that the interpretation of ANOVA Type-I can be altered by simply rearranging the model specification (e.g., from $y \sim m_1 + m_2$ to $y \sim m_2 + m_1$ if $m_1$ and $m_2$ are correlated). Despite posing a threat to the validity of previous work's conclusion, little is known about whether feature selection techniques mitigate correlated metrics in defect datasets.

**Approach**. To identify correlated metrics, we apply correlation analyses. In this paper, we focus on two types of correlation among metrics, i.e., collinearity and multicollinearity. Collinearity is a phenomenon in which one metric can be linearly predicted by another metric. On the other hand, multicollinearity is a phenomenon in which one metric can be linearly predicted by a combination of two or more metrics. We describe each step below.

*(Step 1) Analyse collinearity.* We start from the produced subsets of metrics from RQ1 (*cf.* the Step 2 of RQ1). To analyse collinearity, we use a Spearman rank correlation test ($\rho$) to measure the correlation between metrics. We choose the

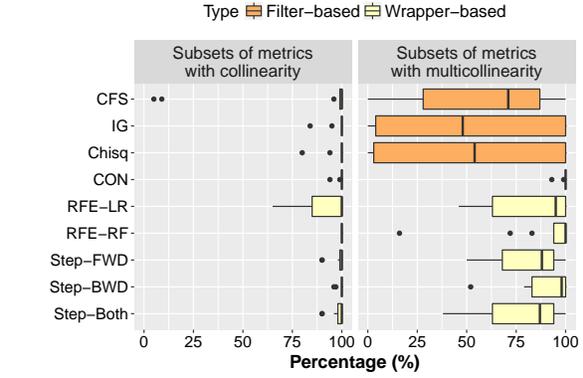

Figure 3: The percentage of subsets of metrics that contain correlated metrics for each studied feature selection technique. The left boxplots present the percentage of subsets of metrics with collinearity, while the right boxplots present the percentage of subsets of metrics with multicollinearity.

Spearman test instead of other correlation tests (e.g., Pearson) since the Spearman test is resilient to non-normal distributions as commonly present in defect datasets. We use the interpretation of correlation coefficients ($|\rho|$) as provided by Kraemer *et al.* [42], i.e., a Spearman correlation coefficient of above *0.7 is considered as a strong correlation*. Thus, two metrics which have their Spearman correlation coefficient above 0.7 are considered correlated. We use the implementation of the `rcorr` function as provided by the `Hmisc` R package [30].

*(Step 2) Analyse multicollinearity.* Similar to Step 1, we start from the produced subsets of metrics from RQ1 (*cf.* the Step 2 of RQ1). To analyse multicollinearity, we use the Variance Inflation Factor analysis (VIF) [20]. VIF determines how well a metric can be linearly predicted by a combination of other metrics through a construction of a regression model. A VIF score of a metric under examination is an $R^2$ goodness-of-fit of the model that is constructed by the other metrics to predict the metric under examination where a VIF score is $\frac{1}{1-R^2}$. We use a VIF threshold value of 5 to identify the presence of multicollinearity, as suggested by Fox [19] and prior work [5, 34, 35, 49]. Thus, metrics that have their VIF score above 5 are considered correlated. We use the implementation of the the Variance Inflation Factor analysis using the `vif` function as provided by the `rms` R package [32]. Finally, we present the results using boxplots in Figure 3.

**Results**. **All of the studied feature selection techniques do not mitigate correlated metrics.** Figure 3 presents the percentage of subsets of metrics that contain correlated metrics for each studied feature selection technique. The studied feature selection techniques produce, at the median, 100% of subsets of metrics with collinearity and 48-100% of subsets of metrics with multicollinearity. Surprisingly, while CFS is a correlation-based feature selection technique, CFS tends to focus on the correlation of each metric and the outcome more than the correlation between metrics. Thus, when CFS is applied, correlated metrics are often selected if these correlated metrics share a strong relationship with the outcome. Since we find that all of the studied feature selection techniques do not mitigate correlated metrics, our findings suggest that correlation analyses should be applied (e.g., the Spearman rank correlation test and the Variance Inflation Factor analysis).

> *Feature selection techniques do not mitigate correlated metrics. The studied feature selection techniques produce up to 100% of subsets of metrics with collinearity and multicollinearity, suggesting that correlation analyses (e.g., a Spearman rank correlation test and a Variance Inflation Factor analysis (VIF)) should be applied to mitigate correlated metrics.*

## V. AN AUTOMATED METRIC SELECTION APPROACH

Since our findings suggest that the commonly-used feature selection techniques should not be used when interpreting defect models and correlation analyses must be applied, we then further investigate if subsets of metrics selected through such correlation analyses increase the consistency and impact the performance of defect models.

Prior work often uses correlation analysis techniques (e.g., a Spearman rank correlation test and a Variance Inflation Factor (VIF) analysis) to mitigate correlated metrics [5, 16, 49, 56, 65, 66, 78–80]. Instead of removing all correlated metrics, prior work often manually selects one metric for each group of correlated metrics [49, 56, 78, 79]—assuming that the selected metric is representative to the group of correlated metrics. However, the manual selection approach often varies based on a domain expert and limits the automation of software analytics in practice [14].

To allow the automation of metric selection, we introduce AutoSpearman, an automated metric selection approach based on the Spearman rank correlation test and the VIF analysis for statistical inference. Unlike manual metric selection by a domain expert, AutoSpearman uses a new criterion for automated metric selection to **select one metric of a group of the highest correlated metrics that shares the least correlation with other metrics that are not in that group**. Unlike commonly-used feature selection techniques which consider the relationship between metrics and the outcome (e.g., wrapper-based feature selection), AutoSpearman is performed in an unsupervised fashion (i.e., outcome labels are not required) to avoid producing optimistically biased performance estimates and interpretation.

Below, we describe AutoSpearman using Algorithm 1, where $S$ is a set of Spearman coefficients for each pair of metrics, $C_S$ is a set of Spearman coefficients that are above a Spearman threshold value ($sp.t$), $V$ is a set of VIF scores of metrics, $C_V$ is a set of VIF scores of metrics that are above a VIF threshold value ($vif.t$), and $M'$ is a set of non-correlated metrics based on the Spearman rank correlation test and the Variance Inflation Factor analysis. The high-level concept of AutoSpearman can be summarised into 2 parts:

*(Part 1) Automatically select non-correlated metrics based on a Spearman rank correlation test.* We first measure the

---

**Algorithm 1:** AutoSpearman

**Input** : $M$ is a set of studied metrics,
$sp.t$ is a threshold value for a Spearman rank correlation test,
$vif.t$ is a threshold value for a Variance Inflation Factor analysis.

**Output:** $M'$ is a set of non-correlated metrics based on a Spearman rank correlation test and a Variance Inflation Factor analysis.

1   $M' = M$
2   $S = Spearman(M, M)$
3   $C_S = \{c(m_i, m_j) \in S | abs(c(m_i, m_j)) \geq sp.t\}$
4   $C_S = sort(C_S)$
5   **for** $c(m_i, m_j)$ *in* $C_S$ **do**
6      $selected.metric = min($
      $mean(abs(Spearman(m_i, M - \{m_i, m_j\})))$,
      $mean(abs(Spearman(m_j, M - \{m_i, m_j\}))))$
7      $removed.metric = \{m_i, m_j\} - selected.metric$
8      $C_S = \{c(m_i, m_j) \in C_S |$
      $m_i \neq removed.metric \wedge m_j \neq removed.metric\}$
9      $M' = M' - removed.metric$
10   **end**
11   **repeat**
12      $V = VIF(M')$
13      $C_V = \{v(m_i) \in V | v(m_i) \geq vif.t\}$
14      $removed.metric = \{m_i |$
      $v(m_i) \in C_V \wedge v(m_i) = max(C_V)$
15      $M' = M' - removed.metric$
16   **until** $|C_V| = 0$;
17   **return** $M'$

---

correlation of all metrics using the Spearman rank correlation test ($\rho$) (*cf.* Line 2). We use the interpretation of correlation coefficients ($|\rho|$) as provided by Kraemer *et al.* [42]—i.e., a Spearman correlation coefficient of above or equal to 0.7 is considered a strong correlation. Thus, we only consider the pairs that have an absolute Spearman correlation coefficient of above or equal to the threshold value ($sp.t$) of 0.7 (*cf.* Line 3).

To automatically select non-correlated metrics based on the Spearman rank correlation test, we start from the pair that has the highest Spearman correlation coefficient (*cf.* Line 4). Since the two correlated metrics under examination can be linearly predicted with each other, one of these two metrics must be removed. Thus, we select the metric that has the lowest average values of the absolute Spearman correlation coefficients of the other metrics that are not included in the pair (*cf.* Line 6). That means the removed metric is the metric in the pair that is not selected (*cf.* Line 7). Since the removed metric may be correlated with the other metrics, we remove any pairs of metrics that are correlated with the removed metric (*cf.* Line 8). Finally, we exclude the removed metric from the set of the remaining metrics ($M'$) (*cf.* Line 9). We repeat this process until all pairs of metrics have their Spearman correlation coefficient of below a threshold value of 0.7 (*cf.* Line 5).

*(Part 2) Automatically select non-correlated metrics based on a Variance Inflation Factor analysis.* We first measure the magnitude of multicollinearity of the remaining metrics ($M'$) using the Variance Inflation Factor analysis (*cf.* Line 12). We use a VIF threshold value ($vif.t$) of 5 to identify the presence of multicollinearity, as suggested by Fox [19] and prior work [5, 34, 35, 49] (*cf.* Line 13).

To automatically remove correlated metrics from the Variance Inflation Factor analysis, we identify the removed metric as the metric that has the highest VIF score (*cf.* Line 14). We then exclude the removed metric from the set of the remaining metrics ($M'$) (*cf.* Line 15). We apply the VIF analysis on the remaining metrics until none of the remaining metrics have their VIF scores above or equal to the threshold value (*cf.* Line 16). Finally, `AutoSpearman` produces a subset of non-correlated metrics based on the Spearman rank correlation test and the VIF analysis ($M'$) (*cf.* Line 17).

## VI. AN EVALUATION OF AUTOSPEARMAN

In this section, we discuss the motivation, approach, and evaluation results with respect to research questions 3 and 4. In particular, we investigate (1) the consistency of subsets of metrics that are produced by `AutoSpearman`; and (2) the impact of `AutoSpearman` on the performance of defect models.

*(RQ3) What is the consistency of subsets of metrics in defect datasets that are produced by* `AutoSpearman`?

**Motivation**. The results of RQ1 show that many commonly-used feature selection techniques often produce low consistency of subsets of metrics among different training samples, yet, little is known about whether `AutoSpearman` yields a higher consistency of subsets of metrics among different training samples compared to the studied feature selection techniques.

**Approach**. To address RQ3, we revisit RQ1 to investigate the consistency of metrics of `AutoSpearman` among different training samples. Similar to RQ1, we follow Step 1 in order to produce training samples, while Step 2 is performed differently. In Step 2, we apply `AutoSpearman` on training samples in order to produce subsets of metrics. We again follow Step 3 of RQ1 in order to analyse the consistency of subsets of metrics among different training samples. Finally, we report the results using boxplots in Figure 4.

**Results**. **`AutoSpearman` yields the highest consistency of subsets of metrics among different training samples when comparing to all studied feature selection techniques.** Figure 4 shows the percentage of metrics that are consistently selected when applying feature selection techniques and `AutoSpearman` to different training samples. When applying `AutoSpearman` to different training samples from the same dataset, we find that, at the median, 69% of the metrics are consistently selected, which leads to improvements of up to 86% from the studied commonly-used feature selection techniques.

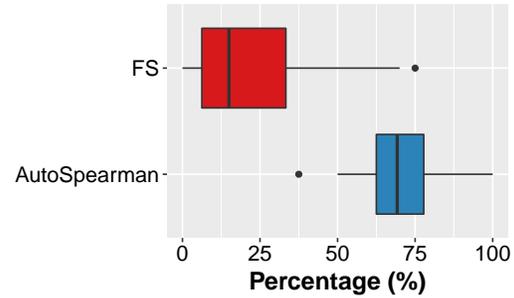

Figure 4: The percentage of metrics that are consistently selected when applying `AutoSpearman` and feature selection techniques to different training samples.

> `AutoSpearman` yields the highest consistency of subsets of metrics among different training samples when comparing to all studied feature selection techniques. When applying `AutoSpearman` to different training samples from the same dataset, we find that, at the median, 69% of the metrics are consistently selected, which leads to improvements of up to 86% from the studied commonly-used feature selection techniques.

*(RQ4) What is the impact of* `AutoSpearman` *on the performance of defect models?*

**Motivation**. Prior research effort has shown the benefits of applying feature selection techniques to defect prediction models [24, 46, 83]. For example, Ghotra *et al.* [24], Lu *et al.* [46], and Xu *et al.* [83] investigate the impact of feature selection techniques on the performance of defect models. Thus, we set out to investigate the impact of `AutoSpearman` on the performance of defect models when comparing to the commonly-used feature selection techniques.

**Approach**. To address RQ4, we analyse the performance of defect models that are constructed using the subsets of metrics that are produced by `AutoSpearman`, the nine studied feature selection techniques, and a baseline (i.e., all metrics of a defect dataset). We describe each step below.

*(Step 1) Generate training samples.* Similar to RQ1, we use the out-of-sample bootstrap validation technique to generate training samples.

*(Step 2) Apply* `AutoSpearman` *and the nine studied feature selection techniques.* Similar to RQ1, we apply `AutoSpearman` and the nine studied feature selection techniques on training samples in order to produce subsets of metrics.

*(Step 3) Construct defect models.* For each training sample, we construct logistic regression [29] and random forest [6] models using subsets of metrics that are produced by `AutoSpearman`, the nine studied feature selection techniques, and all metrics of a defect dataset. We use the implementation of logistic regression as provided by the `glm` function of the `stats` R package [77] with the default parameter setting. We use the implementation of random forest as provided by the `randomForest` function of the `randomForest` R package [7] with the default `ntree` value of 100, since recent studies [73, 76] show that the performance of random forest models is insensitive to the parameter setting. To ensure that the training and testing corpora share similar characteristics and are representative to the original dataset, we do not re-balance nor do we re-sample the training data [71].

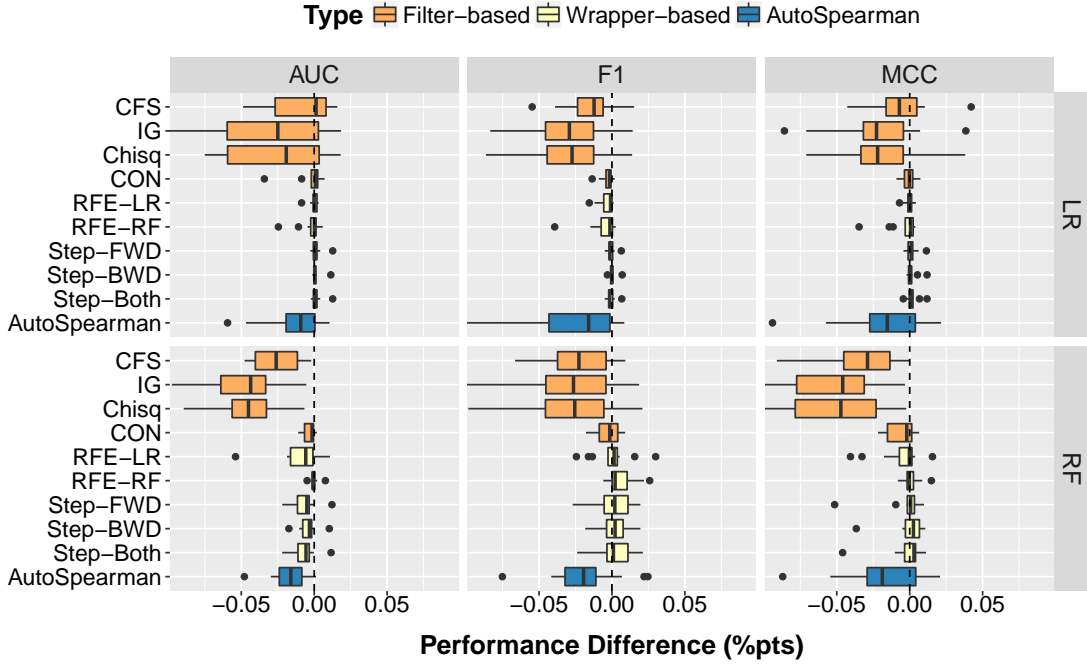

Figure 5: The distributions of the performance difference (%pts) between defect models that are constructed using subsets of metrics that are produced by AutoSpearman, the nine studied feature selection techniques, and all metrics of a defect dataset, i.e., $P_{\text{AutoSpearman,FS}} - P_{\text{All}}$.

*(Step 4) Evaluate defect models.* In our study, we evaluate defect models using three performance measures. First, we use the Area Under the receiver operator characteristic Curve (AUC) to measure the discriminatory power of our models, as suggested by recent research [23, 44, 60, 70]. The *AUC* is a threshold-independent performance measure that evaluates the ability of classifiers in discriminating between defective and clean modules. The values of AUC range between 0 (worst performance), 0.5 (no better than random guessing), and 1 (best performance) [28]. Second, we use the F-measure, i.e, a threshold-dependent measure. F-measure is a harmonic mean (i.e., $\frac{2 \cdot \text{precision} \cdot \text{recall}}{\text{precision} + \text{recall}}$) of precision ($\frac{\text{TP}}{\text{TP+FP}}$) and recall ($\frac{\text{TP}}{\text{TP+FN}}$). Similar to prior studies [3, 86], we use the default probability value of 0.5 as a threshold value for the confusion matrix, i.e., if a module has a predicted probability above 0.5, it is considered defective; otherwise, the module is considered clean. Third, we use the Matthews Correlation Coefficient (MCC) measure, i.e, a threshold-dependent measure, as suggested by prior studies [47, 63]. MCC is a balanced measure based on true and false positives and negatives that is computed using the following equation: $\frac{\text{TP} \times \text{TN} - \text{FP} \times \text{FN}}{\sqrt{(\text{TP+FP})(\text{TP+FN})(\text{TN+FP})(\text{TN+FN})}}$.

*(Step 5) Analyse the impact on model performance.* We analyse the performance difference between defect models that are constructed using subsets of metrics that are produced by feature selection techniques (i.e., AutoSpearman and the nine studied feature selection techniques) and all metrics of a defect dataset ($P_{\text{AutoSpearman,FS}} - P_{\text{All}}$).

Finally, we present the results using boxplots in Figure 5.

**Results**. **AutoSpearman only impacts the performance of defect models by 1-2%pts for AUC, F-measure, and MCC measures, respectively.** Figure 5 shows the performance difference (%pts) between defect models that are constructed using subsets of metrics that are produced by AutoSpearman, the nine studied feature selection techniques, and all metrics of a defect dataset, i.e., $P_{\text{AutoSpearman,FS}} - P_{\text{All}}$. We observe that AutoSpearman only impacts the performance of defect models by 1%pts, 2%pts, and 2%pts (at the median) for AUC, F-measure, and MCC measures, respectively. We also observe that wrapper-based feature selection techniques have the least impact on model performance for both classification techniques. This finding is consistent with Ghotra *et al.* [24] who find that the performance of defect models is impacted by at most 2%pts for the AUC measure when applying wrapper-based feature selection techniques. We observe that filter-based feature selection techniques (except for consistency-based feature selection techniques) have the highest impact on the performance of defect models, since filter-based techniques tend to overly remove metrics that share a strong relationship with the outcome.

> `AutoSpearman` only impacts the performance of defect models by 1-2%pts for AUC, F-measure, and MCC measures, respectively.

## VII. THREATS TO VALIDITY

Like any empirical study design, experimental design settings may impact the results of our study [69]. We now discuss threats to the validity of our study.

### A. Construct Validity

Plenty of prior work show that the parameters of classification techniques have an impact on the performance of defect models [22, 41, 50, 51, 73]. While we use a default `ntree` value of 100 for random forest models, recent studies [33, 73, 81] show that the performance of random forest models is insensitive to the parameter setting. Thus, we believe that this threat is not a major limitation of our work.

The concept of non-correlated metrics in our paper relies on threshold values of correlation analyses (i.e., 0.7 for a Spearman rank correlation test and 5 for a Variance Inflation Factor analysis). Thus, an in-depth sensitivity analysis of these thresholds will be included in future work.

### B. Internal Validity

We studied a limited number of feature selection techniques. Thus, our results may not generalise to other feature selection techniques. Nonetheless, other feature selection techniques can be explored in future work. We provide a detailed methodology for others who would like to re-examine our findings using unexplored feature selection techniques.

### C. External Validity

The studied defect datasets are part of several systems (e.g., Eclipse) that span both proprietary and open source domains. However, we studied a limited number of defect datasets. Thus, the results may not generalise to other datasets and domains. Additional replication studies are needed.

The conclusions of our case study rely on one defect prediction scenario (i.e., within-project defect models). However, there are a variety of defect prediction scenarios in the literature (e.g., cross-project defect prediction [9, 86], just-in-time defect prediction [37], heterogenous defect prediction [57]). Therefore, the conclusions may differ for other scenarios. Thus, future research should revisit our study in other scenarios of defect models.

## VIII. CONCLUSIONS

The interpretation of defect models heavily relies on software metrics that are used to construct them. However, such software metrics are often correlated in defect models. Prior work often uses feature selection techniques to remove correlated metrics in order to improve the performance of defect models. Yet, the interpretation of defect models may be misleading if feature selection techniques produce subsets of inconsistent and correlated metrics.

In this paper, we set out to investigate the consistency and correlation of subsets of metrics that are produced by feature selection techniques. Through a case study of 13 publicly-available defect datasets of systems that span both proprietary and open source domains, we record the following observations:

– When applying the studied feature selection techniques to different training samples from the same dataset, only 6-41% of the metrics are consistently selected. On the other hand, when applying the studied feature selection techniques to the training sample, only 0-15% of the metrics are consistently selected.
– Up to 100% of the subsets of metrics produced by the studied feature selection techniques contain correlated metrics.
– `AutoSpearman` yields the highest consistency of subsets of metrics among different training samples when comparing to all studied feature selection techniques.
– `AutoSpearman` only impacts the performance of defect models by 1-2%pts for AUC, F-measure, and MCC measures, respectively.

Finally, we would like to emphasise that the goal of this work is not to claim the generalisation of our results for every dataset and every model in software engineering. In addition, the best subset of metrics that one should include in studies depends on the goal of the studies. For example, if the goal of the study is *prediction* (i.e., aiming to achieve the highest predictive performance), one might prioritise resources on improving the model performance using feature selection techniques [3, 8, 12, 18, 38, 54, 58, 67] or dimensionality reduction techniques [15] regardless of correlation among metrics. On the other hand, if the goal of the study is *interpretation* (i.e., aiming to examine the impact of various phenomena on software quality), one should avoid using commonly-used feature selection techniques to mitigate correlated metrics. Thus, to automatically mitigate correlated metrics when interpreting defect models, we recommend future studies use `AutoSpearman` in lieu of commonly-used feature selection techniques. Finally, we provide an implementation of `AutoSpearman` as an R package [2].

## ACKNOWLEDGEMENTS

This study would not have been possible without the data shared in the Tera-PROMISE repository [53], as well as the data shared by Shepperd *et al.* [64], Kim *et al.* [39, 82], D'Ambros *et al.* [11, 12], Zimmermann *et al.* [87], as well as supercomputing resources provided by the Phoenix HPC service at the University of Adelaide. This work was supported by the University of Adelaide's Beacon of Enlightenment PhD scholarship and the Australian Research Council's Discovery Early Career Researcher Award (DECRA) funding scheme (DE180100153).

## REFERENCES

[1] "Online Appendix for "AutoSpearman: Automatically Mitigating Correlated Software Metrics for Interpreting Defect Models"," https://github.com/software-analytics/autospearman-appendix.

[2] "Rnalytica: An R package of the Miscellaneous Functions for Data Analytics Research," https://github.com/software-analytics/Rnalytica.

[3] E. Arisholm, L. C. Briand, and E. B. Johannessen, "A Systematic and Comprehensive Investigation of Methods to Build and Evaluate Fault Prediction Models," *Journal of Systems and Software*, vol. 83, no. 1, pp. 2–17, 2010.

[4] W. D. Berry, *Understanding Regression Assumptions*. Sage Publications, 1993, vol. 92.

[5] N. Bettenburg and A. E. Hassan, "Studying the Impact of Social Structures on Software Quality," in *Proceedings of the International Conference on Program Comprehension (ICPC)*, 2010, pp. 124–133.

[6] L. Breiman, "Random forests," *Machine learning*, vol. 45, no. 1, pp. 5–32, 2001.

[7] L. Breiman, A. Cutler, A. Liaw, and M. Wiener, "randomForest : Breiman and Cutler's Random Forests for Classification and Regression. R package version 4.6-12." *Software available at URL: https://cran.r-project.org/web/packages/randomForest*, 2006.

[8] J. Cahill, J. M. Hogan, and R. Thomas, "Predicting Fault-prone Software Modules with Rank Sum Classification," in *Proceedings of the Australian Software Engineering Conference (ASWEC)*, 2013, pp. 211–219.

[9] G. Canfora, A. De Lucia, M. Di Penta, R. Oliveto, A. Panichella, and S. Panichella, "Multi-objective Cross-project Defect Prediction," in *Proceedings of the International Conference on Software Testing, Verification and Validation (ICST)*, 2013, pp. 252–261.

[10] J. M. Chambers, "Statistical Models in S. Wadsworth," *Pacific Grove, California*, 1992.

[11] M. D'Ambros, M. Lanza, and R. Robbes, "An Extensive Comparison of Bug Prediction Approaches," in *Proceedings of the International Conference on Mining Software Repositories (MSR)*, 2010, pp. 31–41.

[12] ——, "Evaluating Defect Prediction Approaches: A Benchmark and an Extensive Comparison," *Empirical Software Engineering (EMSE)*, vol. 17, no. 4-5, pp. 531–577, 2012.

[13] M. Dash, H. Liu, and H. Motoda, "Consistency based Feature Selection," in *Proceedings of the Pacific-Asia Conference on Knowledge Discovery and Data Mining (PAKDD)*, 2000, pp. 98–109.

[14] G. B. de Pádua and W. Shang, "Studying the relationship between exception handling practices and post-release defects," in *Proceedings of the International Conference on Mining Software Repositories (MSR)*, 2018, pp. 564–575.

[15] G. Denaro and M. Pezzè, "An empirical evaluation of fault-proneness models," in *Proceedings of the 24th International Conference on Software Engineering (ICSE)*, 2002, pp. 241–251.

[16] P. Devanbu, T. Zimmermann, and C. Bird, "Belief & Evidence in Empirical Software Engineering," in *Proceedings of the International Conference on Software Engineering (ICSE)*, 2016, pp. 108–119.

[17] B. Efron and R. J. Tibshirani, *An Introduction to the Bootstrap*. Boston, MA: Springer US, 1993.

[18] K. O. Elish and M. O. Elish, "Predicting Defect-prone Software Modules using Support Vector Machines," *Journal of Systems and Software*, vol. 81, no. 5, pp. 649–660, 2008.

[19] J. Fox, *Applied regression analysis and generalized linear models*. Sage Publications, 2015.

[20] J. Fox and G. Monette, "Generalized Collinearity Diagnostics," *Journal of the American Statistical Association (JASA)*, vol. 87, no. 417, pp. 178–183, 1992.

[21] J. Friedman, T. Hastie, and R. Tibshirani, *The Elements of Statistical Learning*. Springer series in statistics, 2001, vol. 1.

[22] W. Fu, T. Menzies, and X. Shen, "Tuning for Software Analytics: Is it really necessary?" *Information and Software Technology*, vol. 76, pp. 135–146, 2016.

[23] B. Ghotra, S. McIntosh, and A. E. Hassan, "Revisiting the Impact of Classification Techniques on the Performance of Defect Prediction Models," in *Proceedings of the International Conference on Software Engineering (ICSE)*, 2015, pp. 789–800.

[24] B. Ghotra, S. Mcintosh, and A. E. Hassan, "A large-scale study of the impact of feature selection techniques on defect classification models," in *Proceedings of the 14th International Conference on Mining Software Repositories*. IEEE Press, 2017, pp. 146–157.

[25] Y. Gil and G. Lalouche, "On the Correlation between Size and Metric Validity," *Empirical Software Engineering (EMSE)*, vol. 22, no. 5, pp. 2585–2611, 2017.

[26] I. Guyon and A. Elisseeff, "An Introduction to Variable and Feature Selection," *Journal of Machine Learning Research*, vol. 3, pp. 1157–1182, 2003.

[27] M. A. Hall, "Correlation-based feature selection for machine learning," Ph.D. dissertation, University of Waikato Hamilton, 1999.

[28] J. a. Hanley and B. J. McNeil, "The meaning and use of the area under a receiver operating characteristic (ROC) curve." *Radiology*, vol. 143, no. 4, pp. 29–36, 1982.

[29] F. E. Harrell, "Ordinal Logistic Regression," in *Regression modeling strategies*. Springer, 2001, pp. 331–343.

[30] F. E. Harrell Jr, "Hmisc: Harrell miscellaneous. R package version 3.12-2," *Software available at URL: http://cran.r-project.org/web/packages/Hmisc*, 2013.

[31] ——, *Regression Modeling Strategies : With Applications to Linear Models, Logistic and Ordinal Regression, and Survival Analysis*. Springer, 2015.

[32] ——, "rms: Regression Modeling Strategies. R package version 5.1-1," 2017.

[33] Y. Jiang, B. Cukic, and T. Menzies, "Can Data Transformation Help in the Detection of Fault-prone Modules?" in *Proceedings of the International Workshop on Defects in Large Software Systems (DEFECTS)*, 2008, pp. 16–20.

[34] J. Jiarpakdee, C. Tantithamthavorn, and A. E. Hassan, "The Impact of Correlated Metrics on Defect Models," *arXiv preprint arXiv:1801.10271*, p. To Appear, 2018.

[35] J. Jiarpakdee, C. Tantithamthavorn, A. Ihara, and K. Matsumoto, "A Study of Redundant Metrics in Defect Prediction Datasets," in *Proceedings of the International Symposium on Software Reliability Engineering Workshops (ISSREW)*, 2016, pp. 51–52.

[36] G. H. John, R. Kohavi, and K. Pfleger, "Irrelevant Features and the Subset Selection Problem," in *Proceedings of the International Conference on Machine Learning (ICML)*, 1994, pp. 121–129.

[37] Y. Kamei, E. Shihab, B. Adams, A. E. Hassan, A. Mockus, A. Sinha, and N. Ubayashi, "A Large-Scale Empirical Study of Just-In-Time Quality Assurance," *Transactions on Software Engineering (TSE)*, vol. 39, no. 6, pp. 757–773, 2013.

[38] A. Kaur and R. Malhotra, "Application of Random Forest in Predicting Fault-prone Classes," in *Proceedings of International Conference on the Advanced Computer Theory and Engineering (ICACTE)*, 2008, pp. 37–43.

[39] S. Kim, H. Zhang, R. Wu, and L. Gong, "Dealing with Noise in Defect Prediction," in *Proceedings of the International Conference on Software Engineering (ICSE)*, 2011, pp. 481–490.

[40] R. Kohavi and G. H. John, "Wrappers for Feature Subset Selection," *Artificial Intelligence*, vol. 97, no. 1-2, pp. 273–324, 1997.

[41] A. G. Koru and H. Liu, "An Investigation of the Effect of Module Size on Defect Prediction Using Static Measures," *Software Engineering Notes (SEN)*, vol. 30, pp. 1–5, 2005.

[42] H. C. Kraemer, G. A. Morgan, N. L. Leech, J. A. Gliner, J. J. Vaske, and R. J. Harmon, "Measures of Clinical Significance," *Journal of the American Academy of Child & Adolescent Psychiatry (JAACAP)*, vol. 42, no. 12, pp. 1524–1529, 2003.

[43] M. Kuhn, J. Wing, S. Weston, A. Williams, C. Keefer, A. Engelhardt, T. Cooper, Z. Mayer, B. Kenkel, R. Team *et al.*, "caret: Classification and regression training. R package version 6.0–78," *Software available at URL: https://cran.r-project.org/web/packages/caret*, 2017.

[44] S. Lessmann, B. Baesens, C. Mues, and S. Pietsch, "Benchmarking Classification Models for Software Defect Prediction: A Proposed Framework and Novel Findings," *Transactions on Software Engineering (TSE)*, vol. 34, no. 4, pp. 485–496, 2008.

[45] J. Li, K. Cheng, S. Wang, F. Morstatter, R. P. Trevino, J. Tang, and H. Liu, "Feature Selection: A Data Perspective," *ACM Computing Surveys (CSUR)*, vol. 50, no. 6, p. 94, 2017.

[46] H. Lu, E. Kocaguneli, and B. Cukic, "Defect Prediction between Software Versions with Active Learning and Dimensionality Reduction," in *Proceedings of the International Symposium on Software Reliability Engineering (ISSRE)*, 2014, pp. 312–322.

[47] B. W. Matthews, "Comparison of the predicted and observed secondary structure of T4 phage lysozyme," *Biochimica et Biophysica Acta (BBA)- Protein Structure*, vol. 405, no. 2, pp. 442–451, 1975.

[48] M. L. McHugh, "The Chi-square Test of Independence," *Biochemia Medica*, vol. 23, no. 2, pp. 143–149, 2013.

[49] S. McIntosh, Y. Kamei, B. Adams, and A. E. Hassan, "The Impact of Code Review Coverage and Code Review Participation on Software Quality," in *Proceedings of the International Conference on Mining Software Repositories (MSR)*, 2014, pp. 192–201.

[50] T. Mende, "Replication of Defect Prediction Studies: Problems, Pitfalls and Recommendations," in *Proceedings of the International Conference on Predictive Models in Software Engineering (PROMISE)*, 2010, pp. 1–10.

[51] T. Mende and R. Koschke, "Revisiting the Evaluation of Defect Prediction Models," *Proceedings of the International Conference on Predictive Models in Software Engineering (PROMISE)*, pp. 7–16, 2009.

[52] T. Menzies, "The Unreasonable Effectiveness of Software Analytics," *IEEE Software*, vol. 35, no. 2, pp. 96–98, March 2018.

[53] T. Menzies, B. Caglayan, E. Kocaguneli, J. Krall, F. Peters, and B. Turhan, "The Promise Repository of Empirical Software Engineering Data," 2012.

[54] T. Menzies, J. Greenwald, and A. Frank, "Data Mining Static Code Attributes to Learn Defect Predictors," *Transactions on Software Engineering (TSE)*, vol. 33, no. 1, pp. 2–13, 2007.

[55] T. M. Mitchell, "Machine Learning," *McGraw Hill*, 1997.

[56] R. Morales, S. McIntosh, and F. Khomh, "Do Code Review Practices Impact Design Quality? : A Case Study of the Qt, VTK, and ITK Projects," in *Proceedings of the International Conference on Software Analysis, Evolution and Reengineering (SANER)*, 2015, pp. 171–180.

[57] J. Nam, W. Fu, S. Kim, T. Menzies, and L. Tan, "Heterogeneous Defect Prediction," *Transactions on Software Engineering (TSE)*, p. In Press, 2017.

[58] A. Okutan and O. T. Yıldız, "Software Defect Prediction using Bayesian Networks," *Empirical Software Engineering (EMSE)*, vol. 19, no. 1, pp. 154–181, 2014.

[59] J. Petrić, D. Bowes, T. Hall, B. Christianson, and N. Baddoo, "The Jinx on the NASA Software Defect Data Sets," in *Proceedings of the International Conference on Evaluation and Assessment in Software Engineering (EASE)*, 2016, pp. 13–17.

[60] F. Rahman and P. Devanbu, "How, and Why, Process Metrics are Better," in *Proceedings of the International Conference on Software Engineering (ICSE)*, 2013, pp. 432–441.

[61] G. Robles, "Replicating MSR: A Study of the Potential Replicability of Papers Published in the Mining Software Repositories Proceedings," in *Proceedings of the International Conference on Mining Software Repositories (MSR)*, 2010, pp. 171–180.

[62] P. Romanski and L. Kotthoff, "FSelector: Selecting attributes. R package version 0.19," *Software available at URL: https://cran.r-project.org/web/packages/FSelector*, 2013.

[63] M. Shepperd, D. Bowes, and T. Hall, "Researcher Bias: The Use of Machine Learning in Software Defect Prediction," *Transactions on Software Engineering (TSE)*, vol. 40, no. 6, pp. 603–616, 2014.

[64] M. Shepperd, Q. Song, Z. Sun, and C. Mair, "Data Quality: Some Comments on the NASA Software Defect Datasets," *Transactions on Software Engineering (TSE)*, vol. 39, no. 9, pp. 1208–1215, 2013.

[65] E. Shihab, C. Bird, and T. Zimmermann, "The Effect of Branching Strategies on Software Quality," in *Proceedings of the International Symposium on Empirical Software Engineering and Measurement (ESEM)*, 2012, pp. 301–310.

[66] E. Shihab, Z. M. Jiang, W. M. Ibrahim, B. Adams, and A. E. Hassan, "Understanding the Impact of Code and Process Metrics on Post-release Defects: A Case Study on the Eclipse Project," in *Proceedings of the International Symposium on Empirical Software Engineering and Measurement (ESEM)*, 2010, pp. 4–10.

[67] S. Shivaji, E. J. Whitehead, R. Akella, and S. Kim, "Reducing Features to Improve Code Change-Based Bug Prediction," *Transactions on Software Engineering (TSE)*, vol. 39, no. 4, pp. 552–569, 2013.

[68] C. Strobl, A.-L. Boulesteix, T. Kneib, T. Augustin, and A. Zeileis, "Conditional Variable Importance for Random Forests," *BMC Bioinformatics*, vol. 9, no. 1, p. 307, 2008.

[69] C. Tantithamthavorn, "Towards a Better Understanding of the Impact of Experimental Components on Defect Prediction Modelling," in *Companion Proceeding of the International Conference on Software Engineering (ICSE)*, 2016, pp. 867–870.

[70] C. Tantithamthavorn and A. E. Hassan, "An Experience Report on Defect Modelling in Practice: Pitfalls and Challenges," in *In Proceedings of the International Conference on Software Engineering: Software Engineering in Practice Track (ICSE-SEIP)*, 2018, pp. 286–295.

[71] C. Tantithamthavorn, A. E. Hassan, and K. Matsumoto, "The Impact of Class Rebalancing Techniques on The Performance and Interpretation of Defect Prediction Models," *arXiv preprint arXiv:1801.10269*, 2018.

[72] C. Tantithamthavorn, S. McIntosh, A. E. Hassan, A. Ihara, and K. Matsumoto, "The Impact of Mislabelling on the Performance and Interpretation of Defect Prediction Models," in *Proceeding of the International Conference on Software Engineering (ICSE)*, 2015, pp. 812–823.

[73] C. Tantithamthavorn, S. McIntosh, A. E. Hassan, and K. Matsumoto, "Automated Parameter Optimization of Classification Techniques for Defect Prediction Models," in *Proceedings of the International Conference on Software Engineering (ICSE)*, 2016, pp. 321–332.

[74] ——, "Comments on "Researcher Bias: The Use of Machine Learning in Software Defect Prediction"," *Transactions on Software Engineering (TSE)*, vol. 42, no. 11, pp. 1092–1094, 2016.

[75] ——, "An Empirical Comparison of Model Validation Techniques for Defect Prediction Models," *Transactions on Software Engineering (TSE)*, vol. 43, no. 1, pp. 1–18, 2017.

[76] ——, "The Impact of Automated Parameter Optimization on Defect Prediction Models," *Transactions on Software Engineering (TSE)*, p. In Press, 2018.

[77] R. C. Team and contributors worldwide, "stats : The R Stats Package. R Package. Version 3.4.0," 2017.

[78] P. Thongtanunam, S. McIntosh, A. E. Hassan, and H. Iida, "Revisiting Code Ownership and its Relationship with Software Quality in the Scope of Modern Code Review," in *Proceedings of the International Conference on Software Engineering (ICSE)*, 2016, pp. 1039–1050.

[79] ——, "Review Participation in Modern Code Review," *Empirical Software Engineering (EMSE)*, vol. 22, no. 2, pp. 768–817, 2017.

[80] Y. Tian, M. Nagappan, D. Lo, and A. E. Hassan, "What Are the Characteristics of High-Rated Apps? A Case Study on Free Android Applications," in *Proceedings of the International Conference on Software Maintenance and Evolution (ICSME)*, 2015, pp. 301–310.

[81] A. Tosun and A. Bener, "Reducing False Alarms in Software Defect Prediction by Decision Threshold Optimization," in *Proceedings of the International Symposium on Empirical Software Engineering and Measurement (ESEM)*, 2009, pp. 477–480.

[82] R. Wu, H. Zhang, S. Kim, and S.-C. Cheung, "Relink: Recovering Links between Bugs and Changes," in *Proceedings of the European Software Engineering Conference and the Symposium on the Foundations of Software Engineering (ESEC/FSE)*, 2011, pp. 15–25.

[83] Z. Xu, J. Liu, Z. Yang, G. An, and X. Jia, "The Impact of Feature Selection on Defect Prediction Performance: An Empirical Comparison," in *Proceedings of the International Symposium on Software Reliability Engineering (ISSRE)*, 2016, pp. 309–320.

[84] L. Yu and H. Liu, "Efficient Feature Selection via Analysis of Relevance and Redundancy," *Journal of Machine Learning Research*, vol. 5, no. Oct, pp. 1205–1224, 2004.

[85] F. Zhang, A. E. Hassan, S. McIntosh, and Y. Zou, "The Use of Summation to Aggregate Software Metrics Hinders the Performance of Defect Prediction Models," *Transactions on Software Engineering (TSE)*, vol. 43, no. 5, pp. 476–491, 2017.

[86] T. Zimmermann, N. Nagappan, H. Gall, E. Giger, and B. Murphy, "Cross-project Defect Prediction," in *Proceedings of the European Software Engineering Conference and the Symposium on the Foundations of Software Engineering (ESEC/FSE)*, 2009, pp. 91–100.

[87] T. Zimmermann, R. Premraj, and A. Zeller, "Predicting Defects for Eclipse," in *Proceedings of the International Workshop on Predictor Models in Software Engineering (PROMISE)*, 2007, pp. 9–19.